\documentclass{nature_mod}

\usepackage{url}
\usepackage{graphicx}
\usepackage{xcolor}
\usepackage{booktabs}
\usepackage{multirow}
\newcommand{\aj}{Astron. J.}
\newcommand{\apj}{Astrophys. J.}
\newcommand{\apjl}{Astrophys. J.}
\newcommand{\apjs}{Astrophys. J.}
\newcommand{\aap}{Astron. Astrophys.}
\newcommand{\mnras}{Mon. Not. R. Astron. Soc.}
\newcommand{\nat}{Nature}
\newcommand{\araa}{Ann. Rev. Astron. Astrophys.}
\newcommand{\pasa}{Pubs. Astron. Soc. Australia}
\newcommand{\ssr}{Space Science Reviews}

\usepackage{amssymb,amsmath,caption}

\newcommand{\frb}{FRB\,121102}

\newcommand{\msun}{M$_\odot$}

\newcommand{\be}{\begin{eqnarray}}
\newcommand{\ee}{\end{eqnarray}}

\newcommand{\DM}{{\rm DM}}

\newcommand{\RMunits}{\rm rad \ m^{-2}}

\long\def\symbolfootnote[#1]#2{\begingroup%
\def\thefootnote{\fnsymbol{footnote}}\footnote[#1]{#2}\endgroup}

\title{An extreme magneto-ionic environment associated with the fast radio burst source \frb}

\author{
D.~Michilli$^{1,2,*}$,
A.~Seymour$^{3,*}$,
J.~W.~T.~Hessels$^{1,2,*}$,
L.~G.~Spitler$^{4}$,
V.~Gajjar$^{5,6,7}$,
A.~M.~Archibald$^{2,1}$,
G.~C.~Bower$^{8}$,
S.~Chatterjee$^{9}$,
J.~M.~Cordes$^{9}$,
K.~Gourdji$^{2}$,
G.~H.~Heald$^{10}$,
V.~M.~Kaspi$^{11}$,
C.~J.~Law$^{12}$,
C.~Sobey$^{13,10}$,
E.~A.~K.~Adams$^{1,14}$,
C.~G.~Bassa$^{1}$,
S.~Bogdanov$^{15}$,
C.~Brinkman$^{16}$,
P.~Demorest$^{17}$,
F.~Fernandez$^{3}$,
G.~Hellbourg$^{12}$,
T.~J.~W.~Lazio$^{18}$,
R.~S.~Lynch$^{19,20}$,
N.~Maddox$^{1}$,
B.~Marcote$^{21}$,
M.~A.~McLaughlin$^{22,20}$,
Z.~Paragi$^{21}$,
S.~M.~Ransom$^{23}$,
P.~Scholz$^{24}$,
A.~P.~V.~Siemion$^{12,25,26}$,
S.~P.~Tendulkar$^{11}$,
P.~Van Rooy$^{27}$,
R.~S.~Wharton$^{4}$,
D.~Whitlow$^{3}$
}

\begin{document}

\maketitle

\begin{affiliations}
 \item ASTRON, Netherlands Institute for Radio Astronomy, Postbus 2, 7990 AA, Dwingeloo, The Netherlands
 \item Anton Pannekoek Institute for Astronomy, University of
    Amsterdam, Science Park 904, 1098 XH Amsterdam, The Netherlands
 \item National Astronomy and Ionosphere Center, Arecibo Observatory, PR 00612, USA
 \item Max-Planck-Institut f\"{u}r Radioastronomie, Auf dem H\"{u}gel 69, D-53121 Bonn, Germany
 \item Space Science Laboratory, 7 Gauss Way, University of California, Berkeley, CA 94710, USA
 \item Xinjiang Astronomical Observatory, CAS, 150 Science 1-Street, Urumqi, Xinjiang 830011, China
 \item Key Laboratory of Radio Astronomy, Chinese Academy of Sciences, Nanjing 210008, China
 \item Academia Sinica Institute of Astronomy and Astrophysics, 645 N. A'ohoku Place, Hilo, HI 96720, USA
 \item Cornell Center for Astrophysics and Planetary Science and Department of Astronomy, Cornell University, Ithaca, NY 14853, USA
 \item CSIRO Astronomy and Space Science, 26 Dick Perry Avenue, Kensington, WA 6151, Australia
 \item Department of Physics and McGill Space Institute, McGill University, 3600 University, Montr\'{e}al, QC H3A 2T8, Canada
 \item Department of Astronomy and Radio Astronomy Lab, University of California, Berkeley, CA 94720, USA
 \item International Centre for Radio Astronomy Research - Curtin University, GPO Box U1987, Perth, WA 6845, Australia
 \item Kapteyn Astronomical Institute, University of Groningen, Postbus 800, 9700 AA, Groningen, The Netherlands
 \item Columbia Astrophysics Laboratory, Columbia University,  New York, NY 10027, USA
 \item Physics Department, University of Vermont, Burlington, VT 05401, USA
 \item National Radio Astronomy Observatory, P.O. Box O, Socorro, NM 87801 USA
 \item Jet Propulsion Laboratory, California Institute of Technology, Pasadena, CA 91109, USA
 \item Green Bank Observatory, PO Box 2, Green Bank, WV 24944, USA
 \item Center for Gravitational Waves and Cosmology, Chestnut Ridge Research Building, Morgantown, WV 26505, USA
 \item Joint Institute for VLBI ERIC, Postbus 2, 7990 AA Dwingeloo, The Netherlands
 \item Department of Physics and Astronomy, West Virginia University, Morgantown, WV 26506, USA
 \item National Radio Astronomy Observatory, Charlottesville, VA 22903, USA
 \item National Research Council of Canada, Herzberg Astronomy and Astrophysics, Dominion Radio Astrophysical Observatory, P.O. Box 248, Penticton, BC V2A 6J9, Canada
 \item Radboud University, Nijmegen, Comeniuslaan 4, 6525 HP Nijmegen, The Netherlands
 \item SETI Institute, 189 N Bernardo Ave \#200, Mountain View, CA 94043, USA
 \item Department of Electrical Engineering and Computer Science, Case Western Reserve University, Cleveland, OH 44106, USA\\
$^{*}$ These authors contributed equally to this work
\end{affiliations}

\bigskip
\bigskip

\begin{abstract}
Fast radio bursts (FRBs) are millisecond-duration, extragalactic radio flashes of unknown physical origin\cite{Lor07,Tho13,Pet16}.  \frb, the only known repeating FRB source\cite{Spi14,Spi16,Pet15b}, has been localized to a star-forming region in a dwarf galaxy\cite{Cha17,Ten17,Bas17} at redshift $z = 0.193$, and is spatially coincident with a compact, persistent radio source\cite{Cha17,Mar17}.  The origin of the bursts, the nature of the persistent source, and the properties of the local environment are still debated.
Here we present bursts that show $\sim$100\% linearly polarized emission at a very high and variable Faraday rotation measure in the source frame: $\text{RM$_{\rm src}$} = +1.46 \times 10^5$\,rad\,m$^{-2}$ and $+1.33 \times 10^5$\,rad\,m$^{-2}$ at epochs separated by 7 months, in addition to narrow ($\lesssim 30$\,$\mu$s) temporal structure.  The large and variable rotation measure demonstrates that \frb\ is in an extreme and dynamic magneto-ionic environment, while the short burst durations argue for a neutron star origin.  Such large rotation measures have, until now, only been observed\cite{Bow03,Mar07} in the vicinities of massive black holes ($M_{\rm BH} \gtrsim 10^4$\,\msun).  Indeed, the properties of the persistent radio source are compatible with those of a low-luminosity, accreting massive black hole\cite{Mar17}.  The bursts may thus come from a neutron star in such an environment.  However, the observed properties may also be explainable in other models, such as a highly magnetized wind nebula\cite{Met17} or supernova remnant\cite{Pir16} surrounding a young neutron star.
\end{abstract}

\clearpage

Using the 305-m  William  E.  Gordon Telescope at the Arecibo Observatory, we detected 16 bursts from \frb\ at radio frequencies from $4.1-4.9$\,GHz (Table~\ref{tab:bursts}).  The data recorder provided complete polarization parameters with 10.24-$\mu$s time resolution.  See Methods and Extended Data Figs.~1-6 for observation and analysis details.

The 4.5-GHz bursts have typical widths $\lesssim 1$\,ms, which are narrower than the 2 to 9-ms bursts previously detected at lower frequencies\cite{Spi16,Sch16}.  In some cases they show multiple components and structure close to the sampling time of the data.  Burst \#6 is particularly striking, with a width of $\lesssim 30$\,$\mu$s, which constrains the size of the emitting region to $\lesssim 10$\,km, modulo geometric and relativistic effects.  Evolution in burst morphology with frequency complicates the determination\cite{Spi16} of dispersion measure ($\text{DM} = \int_{0}^{d} n_e(l)~dl$), but aligning the narrow component in Burst \#6 results in DM$ = 559.7 \pm 0.1$\,pc\,cm$^{-3}$, which is consistent\cite{Spi14,Spi16,Sch16,Sch17} with other bursts detected since 2012, and suggests that any {\it bona fide} dispersion measure variations are at the $\lesssim 1$\% level.

\begin{figure}
\centerline{\includegraphics{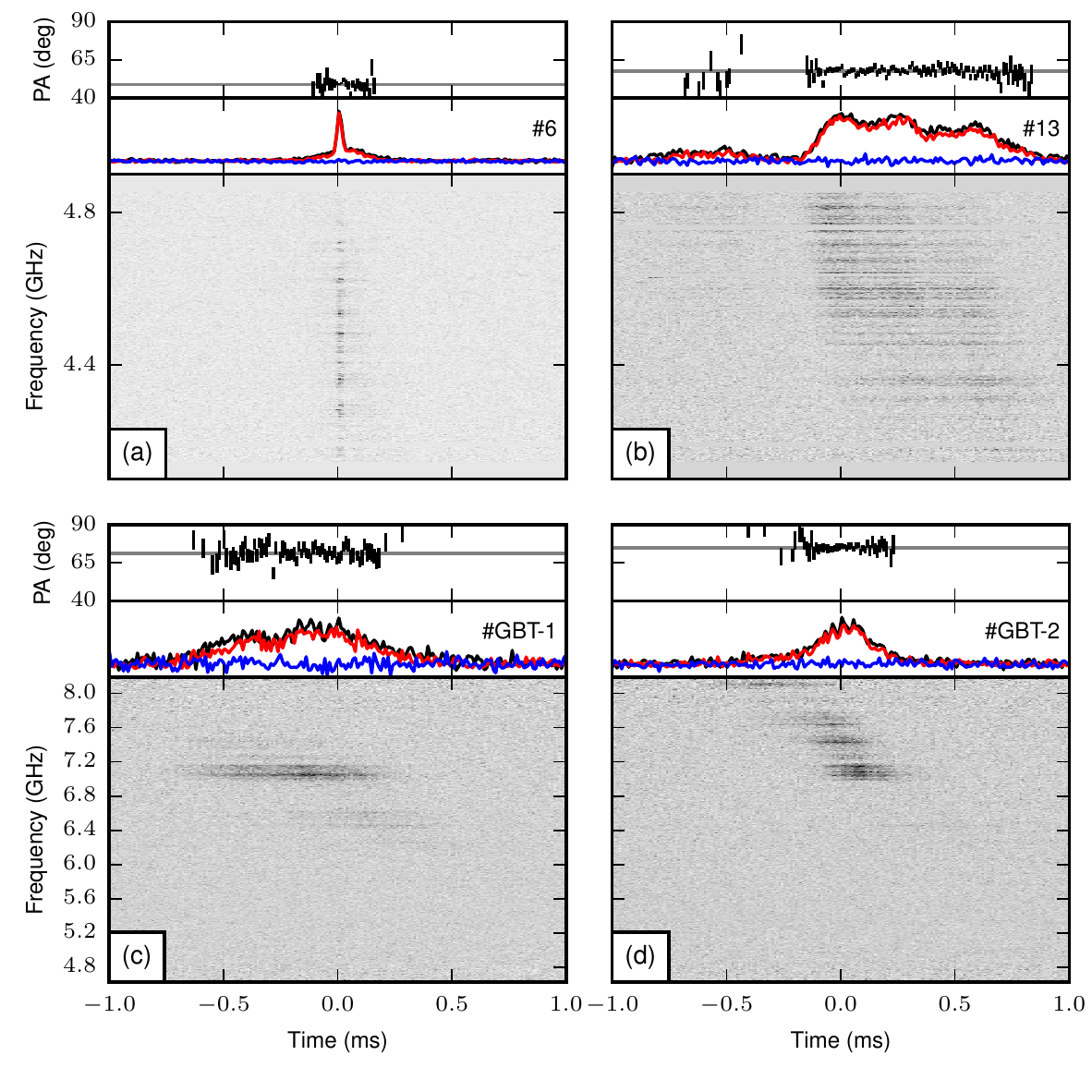}}
\caption{{\bf
Polarization angles, pulse profile and spectrum of four bursts.}
A grey, horizontal line indicates the average PA of each burst.
The red and blue lines indicate linear and circular polarization profiles, respectively, while the black line is the total intensity.
Burst numbers are indicated next to the pulse profiles.
Arecibo bursts (\textbf{a} and \textbf{b}) are plotted with time and frequency resolutions of $10.24$\,$\mu$s and $1.56$\,MHz, respectively.
GBT bursts (\textbf{c} and \textbf{d}) are plotted with time and frequency resolutions of $10.24$\,$\mu$s and $5.86$\,MHz, respectively.
\label{fig:bursts_zoom}}
\end{figure}

\begin{figure}
\centerline{\includegraphics{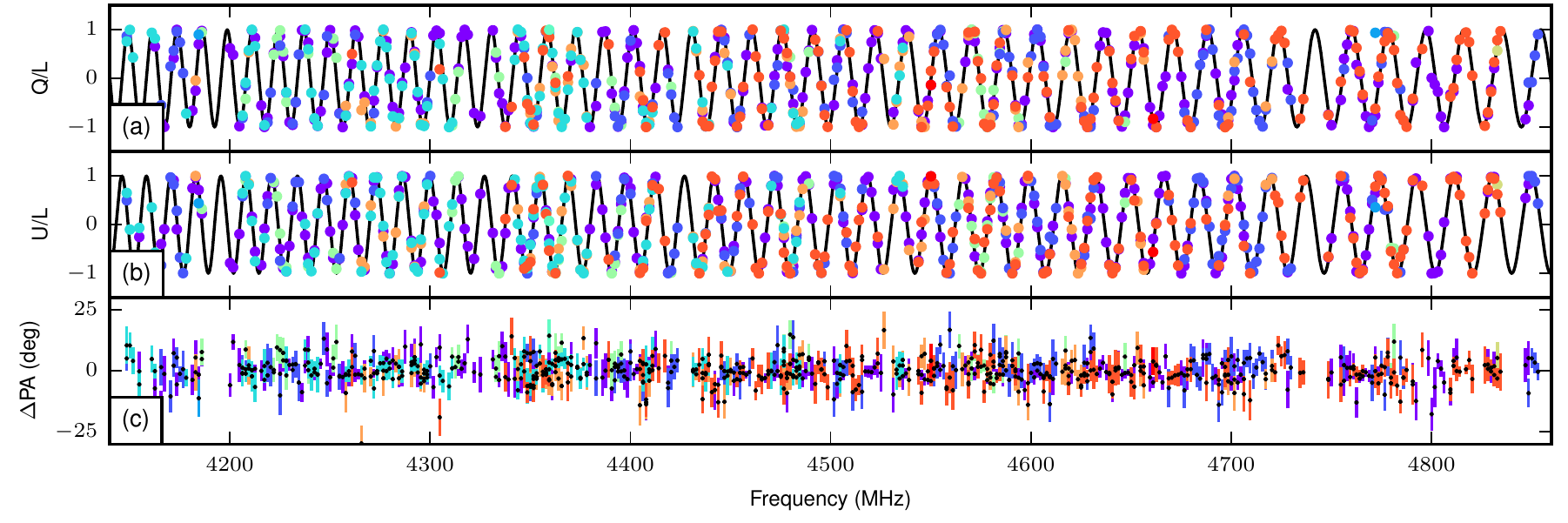}}
\caption{{\bf Faraday rotation in the bursts.}
\textbf{a} and \textbf{b}: variations of the Stokes $Q$ and $U$ parameters with frequency, normalized by the total linear polarization ($L = \sqrt[]{Q^2+U^2}$), for the six brightest Arecibo bursts detected on MJD~57747.
Different bursts are plotted using different colours.
Only data points with S/N $> 5$ are plotted, and do not include uncertainties.
A black line represents the best-fit Faraday rotation model for the global values reported in Table~\ref{tab:bursts}.
\textbf{c}: difference between model and measured PA values with 1-$\sigma$ uncertainties around the central values indicated with black dots.
\label{fig:UQ}}
\end{figure}

After correcting for Faraday rotation, and accounting for $\sim$2\% depolarization from the finite channel widths, the bursts are consistently $\sim$100\% linearly polarized  (Fig.~\ref{fig:bursts_zoom}).  The polarization angles PA = PA$_{\infty}+\theta$ (where PA$_{\infty}$ is a reference angle at infinite frequency, $\theta = {\rm RM}\lambda^2$ is the rotation angle of the electric field vector and $\lambda$ is the observing wavelength) are flat across the observed frequency range and burst envelopes ($\Delta{\rm PA} \lesssim 5^{\circ}$\,ms$^{-1}$).  This could mean that the burst durations reflect the timescale of the emission process and not the rate of a rotating beam sweeping across the line of sight.  Any circular polarization is less than a few percent of the total intensity.  Faraday rotation measure is defined $\text{RM} = 0.81 \int_{d}^{0} B_{\parallel}(l) \cdot n_e(l) ~dl$, where $B_{\parallel}$ is the line-of-sight magnetic field strength ($\mu$G), $l$ is the distance (parsecs), and $n_e$ is the electron density (cm$^{-3}$); by convention RM is positive when the magnetic field points toward the observer.  On average, the observed $\text{RM}_\text{obs} = (+1.027 \pm 0.001) \times 10^5$\,rad\,m$^{-2}$ and varies by $\sim 0.5$\% between Arecibo observing sessions spanning a month (Fig.~\ref{fig:UQ}; Table~\ref{tab:bursts}).  The lack of polarization in previous burst detections\cite{Sch16,Sch17} at $1.1-2.4$\,GHz is consistent with the relatively coarse frequency channels causing bandwidth depolarization and constrains $|\text{RM}_\text{obs}| \gtrsim 10^4$\,rad\,m$^{-2}$ at those epochs.

Confirmation of this extreme Faraday rotation comes from independent observations at 4--8\,GHz with the 110-m Robert C. Byrd Green Bank Telescope (GBT), which yield $\text{RM}_\text{obs} = (+0.935 \pm 0.001) \times 10^5$\,rad\,m$^{-2}$ at an epoch 7 months later than the Arecibo detections.  The GBT and Arecibo $\text{RM}_\text{obs}$ differ with high statistical significance and indicate that the rotation measure can vary by at least 10\% on half-year timescales (Table~\ref{tab:bursts} and Extended Data Fig.~\ref{fig:RMevolution}).

The Faraday rotation must come almost exclusively from within the host galaxy: the expected Milky Way contribution\cite{Opp15} is $-25\pm 80$\,rad\,m$^{-2}$, while estimated intergalactic medium contributions\cite{Aka16} are $\lesssim 10^2$\,rad\,m$^{-2}$.  In the source reference frame, $\text{RM}_\text{src} = \text{RM}_\text{obs} \times (1+z)^2 = +1.46 \times 10^5$\,rad\,m$^{-2}$ and $+1.33 \times 10^5$\,rad\,m$^{-2}$ in the Arecibo and GBT data, respectively.  The observed variations in rotation measure, without a correspondingly large change in dispersion measure, imply that the Faraday rotation comes from a spatially compact region with a high magnetic field. Furthermore, that region must be close to \frb, since it is extremely unlikely that the line of sight coincidentally encounters a small but un-associated structure with the required high magnetic field.

We can fit all 16 Arecibo bursts with a single PA$_{\infty}^{\rm global} = 58^{\circ} \pm 1^{\circ}$ (referenced to infinite frequency; measured counter-clockwise from North to East) and a single RM$_{\rm global}$ per observation day (Table~\ref{tab:bursts}); however, we cannot rule out small changes in the rotation measure ($\lesssim 50$\,rad\,m$^{-2}$) and polarization angle ($\lesssim 10^{\circ}$) between bursts.  The GBT data are in tension with a single PA$_{\infty}^{\rm global}$, but this could be an instrumental difference or reflection of the higher observing frequency.  The near constancy of polarization angle suggests that the burst emitter has a stable geometric orientation with respect to the observer.  The $\gtrsim 98$\% linear polarization fraction at a single RM constrains turbulent scatter\cite{O'Sul12} $\sigma_{RM} < 25$\,rad\,m$^{-2}$ and a linear gradient across the source $\Delta_{RM} < 20$\,rad\,m$^{-2}$, and there is no evidence for deviations from the wavelength-squared ($\lambda^2$) scaling of the Faraday rotation effect.
A Rotation Measure Synthesis and {\tt RMCLEAN} analysis also implies a Faraday thin medium (see Methods).

In the rest frame, the host galaxy contributes $\text{DM}_{\rm Host} \sim 70$--270\,pc\,cm$^{-3}$ to the total dispersion measure of the bursts\cite{Ten17}.  Given RM$_{\rm src}$, this corresponds to an estimated line-of-sight magnetic field $B_{\parallel}^{\rm FRB} = (0.6-2.4) \times f_{\rm DM}$\,mG.  This is a lower limit range because the dispersion measure contribution specifically related to the observed rotation measure (i.e. $\text{DM}_{\rm RM}$) could be much smaller than the total dispersion measure contribution of the host ($\text{DM}_{\rm Host}$, dominated by the star-forming region), which we quantify by the scaling factor $f_{\rm DM} = \text{DM}_{\rm Host} / \text{DM}_{\rm RM} \geq 1$.  For comparison, typical magnetic field strengths within the interstellar medium of our Galaxy\cite{Hav15} are only $\sim$5\,$\mu$G.

We can constrain the electron density ($n_e$), electron temperature ($T_e$), and length scale ($L_{\rm RM}$) of the region causing the Faraday rotation by balancing the magnetic field and thermal energy densities (Extended Data Fig.~\ref{fig:dmrm}).  For example, assuming equipartition and $T_e=10^6$\,K, we find a density of $n_e\sim 10^2 {\rm cm^{-3}}$ on a length scale $L_{\rm RM} \sim 1$\,pc, i.e., comparable to the upper limit on the size of the persistent source\cite{Mar17}.

A star-forming region, like that hosting \frb, will contain HII regions of ionized hydrogen.  While ultracompact HII regions have sufficiently high magnetic fields and electron densities to explain the large rotation measure, the constraints from $\text{DM}_{\rm Host}$ and the absence of free-free absorption of the bursts exclude a wide range of HII region sizes and densities\cite{Hunt09} for typical temperatures of $10^4$\,K.

\begin{figure}
\centerline{\includegraphics{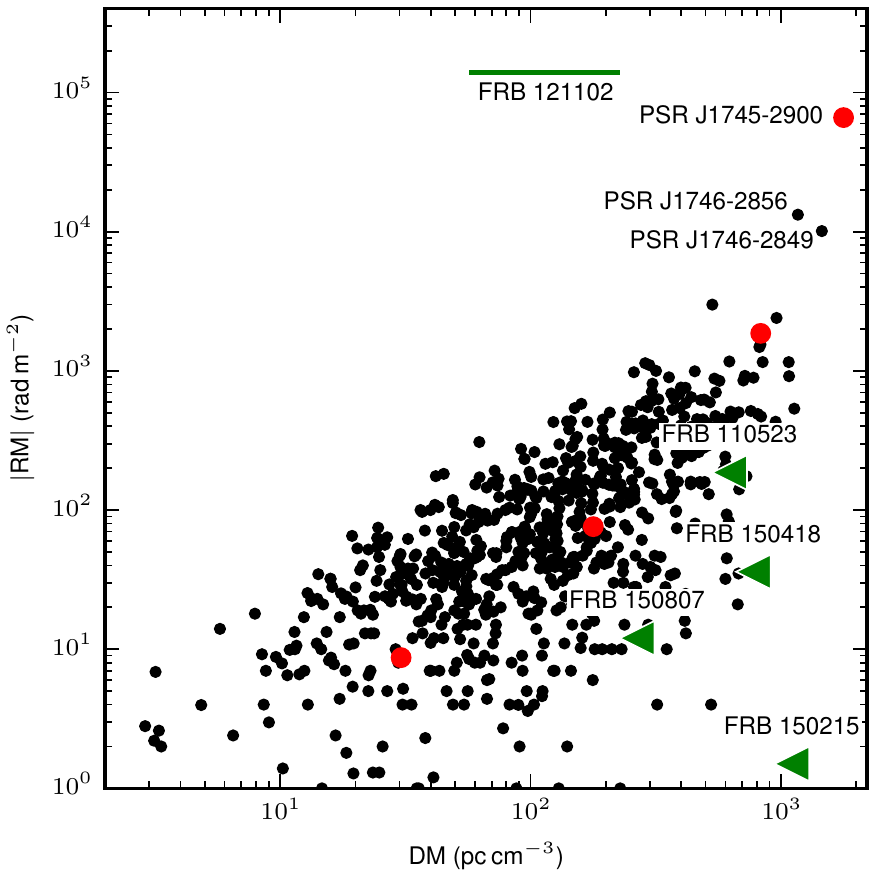}}
\caption{{\bf
Magnitude of rotation measure versus dispersion measure for FRBs and Galactic pulsars.}
Radio-loud magnetars are highlighted with red dots, while radio pulsars and magnetars closest to the Galactic Centre are labelled by name (source: the ATNF Pulsar Catalogue\cite{Man05}).
A green bar represents \frb\ and the uncertainty on the DM contribution of the host galaxy\cite{Ten17}.
Green triangles are other FRBs with measured RM; here the DM is an upper limit on the contribution from the host galaxy.
\label{fig:DM_RM_all}}
\end{figure}

The environment around a massive black hole is consistent with the ($n_e$, $L_{\rm RM}$, $T_e$) constraints\cite{Qua99}, and the properties of the persistent source are compatible with those of a low-luminosity, accreting massive black hole\cite{Mar17}.  The high rotation measure toward the Galactic Centre magnetar\cite{Eat13}
PSR~J1745$-$2900 (Fig.~\ref{fig:DM_RM_all}), $\text{RM} = -7 \times 10^4$\,rad\,m$^{-2}$, provides an intriguing observational analogy for a scenario in which the bursts are produced by a neutron star in the immediate environment of a massive black hole.  However, \frb's bursts are many orders of magnitude more energetic than those of PSR~J1745$-$2900 or any Galactic pulsar.

Alternatively, a millisecond magnetar model has previously been proposed\cite{Ten17,Mar17,Met17} for \frb, and in that model one would expect a surrounding supernova remnant and nebula powered by the central neutron star.  The ($n_e$, $L_{\rm RM}$, $T_e$) constraints are broadly compatible with the conditions in pulsar wind nebulae, but dense filaments like those seen in the Crab Nebula\cite{Dav85} may need to be invoked to explain the high and variable rotation measure of \frb.  In a young neutron star scenario, an expanding supernova remnant could also in principle produce a high rotation measure by sweeping up surrounding ambient medium and progenitor ejecta\cite{Har10}.

A more detailed discussion of these scenarios is provided in the Methods, and more exotic models also remain possible\cite{Zha17}.

Regardless of its nature, \frb\ clearly inhabits an extreme magneto-ionic environment.  In contrast, Galactic pulsars with comparable dispersion measures have rotation measures that are less than a hundredth that of \frb\ (Fig.~\ref{fig:DM_RM_all}).  \frb's ${\rm RM}_{\rm src}$ is also $\sim$500$\times$ larger than that detected in any FRB to date\cite{Mas15}.  The five other known FRBs with polarimetric measurements present a heterogeneous picture, with a range of polarization fractions and rotation measures\cite{Pet16}.  As also previously considered\cite{Pet15a}, \frb\ suggests that FRBs with no detectable linear polarization may actually have very large $|{\rm RM}| \gtrsim 10^4 - 10^5$\,rad\,m$^{-2}$ that was undetectable given the limited frequency resolution (0.4-MHz channels at 1.4\,GHz) of the observations.

Monitoring of \frb's rotation measure and polarization angle with time, along with searches for polarization and Faraday rotation from the persistent source, can help differentiate among models.  \frb\ is peculiar not only because of its large rotation measure but also because it is the only known repeating FRB.  While this may indicate that \frb\ is a fundamentally different type of source compared to the rest of the FRB population, future measurements can investigate a possible correlation between FRB repetition and rotation measure.  Perhaps the markedly higher activity level of \frb\ compared to other known FRBs is predominantly a consequence of its environment; e.g., because these magnetized structures can also boost the detectability of the bursts via plasma lensing\cite{Cor17}.\\

\


\begin{addendum}

\item We heartily thank the staff of both the Arecibo Observatory and Green Bank Observatory for their continued support and dedication to enabling observations like those presented here.  We also thank B.~Adebahr, L.~Connor, G.~Desvignes, R.~Eatough, R.~Fender, M.~Haverkorn, A.~Karastergiou, R.~Morganti, E.~Petroff, F.~Vieyro, and J.~Weisberg for helpful suggestions and comments on the manuscript.  The Arecibo Observatory is operated by SRI International under a cooperative agreement with the National Science Foundation (AST-1100968), and in alliance with Ana G.~M\'{e}ndez-Universidad Metropolitana, and the Universities Space Research Association.  The Green Bank Observatory is a facility of the National Science Foundation operated under cooperative agreement by Associated Universities, Inc.  Breakthrough Listen (BL) is managed by the Breakthrough Initiatives, sponsored by the Breakthrough Prize Foundation (www.breakthroughinitiatives.org).  The research leading to these results has received funding from the European Research Council (ERC) under the European Union's Seventh Framework Programme (FP7/2007-2013).  J.W.T.H. is a Netherlands Organisation for Scientific Research (NWO) Vidi Fellow, and together with D.M., K.G. and C.G.B. also gratefully acknowledges funding for this work from ERC Starting Grant DRAGNET under contract no. 337062.  L.G.S. gratefully acknowledges financial support from the ERC Starting Grant BEACON, under contract number 279702, as well as the Max Planck Society.  A.M.A. is an NWO Veni Fellow.  S.C., J.M.C., P.D., T.J.L., M.A.M., and S.M.R. are partially supported by the NANOGrav Physics Frontiers Center (NSF award 1430284).  V.M.K. holds the Lorne Trottier Chair in Astrophysics \& Cosmology and a Canada Research Chair and receives support from an NSERC Discovery Grant and Herzberg Prize, from an R. Howard Webster Foundation Fellowship from the Canadian Institute for Advanced Research (CIFAR), and from the FRQNT Centre de Recherche en Astrophysique du Qu\'{e}bec.  C.J.L. acknowledges support from NSF award 1611606.
Part of this research was carried out at the Jet Propulsion Laboratory, California Institute of Technology, under a contract with the National Aeronautics and Space Administration.
B.M. acknowledges support from the Spanish Ministerio de Econom\'ia y Competitividad (MINECO) under grants AYA2016-76012-C3-1-P and MDM-2014-0369 of ICCUB (Unidad de Excelencia ``Mar\'ia de Maeztu'').  S.M.R. is a CIFAR Senior Fellow.  P.S. holds a Covington Fellowship at DRAO.

\item[Author Contributions] A.S. led development of the Arecibo observing functionality used here, and discovered the first $\sim 4.5$\,GHz bursts.  L.G.S. is PI of the Arecibo monitoring campaign.  D.M. discovered the rotation measure and analyzed the burst properties in detail.  K.G. comprehensively searched all Arecibo $\sim$4.5\,GHz data sets for bursts.  J.W.T.H. led the discussion of interpretation and writing of the manuscript.  A.M.A. guided development of the RM fitting code. G.H.H. and C.S. performed the RM Synthesis and deconvolution analysis.  G.B., S.C., J.M.C., V.G., V.M.K., C.J.L, M.A.M. and D.M., also made significant contributions to the writing of the manuscript and analysis.  V.G. observed, searched and detected bursts from the GBT at 6.5\,GHz as a part of the BL monitoring campaign of known FRBs. A.Si. is the PI of the BL project. C.B. helped with the polarization calibration of the test pulsar. G.H. wrote a code to splice raw voltages across compute nodes.  All other co-authors contributed significantly to the interpretation of the analysis results and to the final version of the manuscript.

\item[Competing Interests] The authors declare that they have no
  competing financial interests.

\item[Correspondence] Correspondence and requests for materials should
  be addressed to J.W.T.H.~(email: J.W.T.Hessels@uva.nl).

\end{addendum}

\clearpage


\begin{table}
\begin{center}
\scriptsize
\caption{\footnotesize {\bf
Properties of Arecibo and GBT bursts.}
MJDs are referenced to infinite frequency at the solar system barycentre; their uncertainties are of the order of the burst widths.
Widths have uncertainties $\sim 10$\,$\mu$s.
Peak flux densities $S$ and fluences $F$ have $\sim 20\%$ fractional uncertainties.
RMs are not corrected for redshift and PAs are referenced to infinite frequency.
Bursts with no individual RM entry (--) were too weak to reliably fit on their own.
The last two columns refer to a global fit of all bursts.  All errors are $1\sigma$; see Methods for observational details.
\label{tab:bursts}
}
\begin{tabular}{lllllllll}
\toprule
Burst & MJD & Width & S    & F        & RM$_{\rm obs}$ & PA$_{\infty}$ & RM$_{\rm global}$ & PA$^{\rm global}_{\infty}$ \\
      &     & (ms)  & (Jy) & (Jy\,ms) & (rad\,m$^{-2}$) & (deg) & (rad\,m$^{-2}$) & (deg) \\
\midrule
 1 & 57747.1295649013 & 0.80 & 0.9 & 0.7   & +102741 $\pm$ 9 & 49 $\pm$ 2  & \multirow{10}{*}{+102708 $\pm$ 4} & \multirow{16}{*}{58 $\pm$ 1} \\
 2 & 57747.1371866766  & 0.85 & 0.3 & 0.2  & +102732 $\pm$ 34 & 55 $\pm$ 9 & \\
 3 & 57747.1462710273 & 0.22 & 0.8 & 0.2  & +102689 $\pm$ 18 & 64 $\pm$ 5 & \\
 4 & 57747.1515739398 & 0.55 & 0.2 & 0.09 & \multicolumn{1}{c}{--} & \multicolumn{1}{c}{--} & \\
 5 & 57747.1544674919 & 0.76 & 0.2 & 0.1 & \multicolumn{1}{c}{--} & \multicolumn{1}{c}{--} & \\
 6 & 57747.1602892954 & 0.03 & 1.8 & 0.05  & +102739 $\pm$ 35 & 49 $\pm$ 9 & \\
 7 & 57747.1603436945 & 0.31 & 0.6 & 0.2  & +102663 $\pm$ 33 & 71 $\pm$ 9 & \\
 8 & 57747.1658277033 & 1.36 & 0.4 & 0.5  & +102668 $\pm$ 18 & 67 $\pm$ 4 & \\
 9 & 57747.1663749941 & 1.92 & 0.2 & 0.3 & \multicolumn{1}{c}{--} & \multicolumn{1}{c}{--} & \\
10 & 57747.1759674338 & 0.98 & 0.2 & 0.2 & \multicolumn{1}{c}{--} & \multicolumn{1}{c}{--} & \\
\cmidrule{1-8}
11 & 57748.1256436428 & 0.95 & 0.1 & 0.1  & \multicolumn{1}{c}{--} & \multicolumn{1}{c}{--} & \multirow{5}{*}{+102521 $\pm$ 4} & \\
12 & 57748.1535244366 & 0.42 & 0.4 & 0.2  & +102508 $\pm$ 35 & 63 $\pm$ 10 & \\
13 & 57748.1552149312 & 0.78 & 0.8 & 0.6  & +102522 $\pm$ 17 & 59 $\pm$ 4 & \\
14 & 57748.1576076618 & 0.15 & 1.2 & 0.2  & +102489 $\pm$ 18 & 67 $\pm$ 5 & \\
15 & 57748.1756968287 & 0.54 & 0.4 & 0.4  & +102492 $\pm$ 37 & 64 $\pm$ 10 & \\
\cmidrule{1-8}
16 & 57772.1290302972 & 0.74 & 0.8 & 0.6  & +103020 $\pm$ 12 & 64 $\pm$ 3 & +103039 $\pm$ 4 & \\
\midrule
GBT-1 & 57991.5801286366 & 0.59 & 0.4 & 0.2 & +93526 $\pm$ 72 & 73 $\pm$ 8 & \multirow{2}{*}{+93573 $\pm$ 24} & \multirow{2}{*}{68 $\pm$ 2} \\
GBT-2 & 57991.5833032369 & 0.27 & 0.9 & 0.2 & +93533 $\pm$ 42 & 71 $\pm$ 4 & \\
\bottomrule
\end{tabular}
\end{center}
\end{table}

\clearpage

\clearpage

\clearpage

\clearpage


\section*{Methods}

The analyses described here were based on the PRESTO\cite{Ran01}, PSRCHIVE\cite{Str12}, and DSPSR\cite{Str11} pulsar software suites, as well as custom-written Python scripts for linking utilities into reduction pipelines, fitting the data, and plotting.

\subsubsection*{Observations and burst search}

\noindent {\it Arecibo} \\

We observed using the Arecibo `C-band' receiver (dual linear receptors), in the frequency range $4.1-4.9$\,GHz, and the Puerto-Rican Ultimate Pulsar Processing Instrument (PUPPI) backend recorder.  The full list of observations is reported in Extended Data Table~\ref{tab:observations}.  We operated PUPPI in its `coherent search' mode, which produced $10.24$\,$\mu$s samples and $512 \times 1.56$\,MHz frequency channels, each coherently dedispersed to $\text{DM} = 557.0$\,pc\,cm$^{-3}$.  Coherent dedispersion within each 1.56-MHz channel means that the intra-channel dispersive smearing is $< 2$\,$\mu$s even if the burst DM is $10$\,pc\,cm$^{-3}$ higher/lower than the fiducial value of $557.0$\,pc\,cm$^{-3}$ used in the PUPPI recording.  The raw PUPPI data also provide auto- and cross-correlations of the two linear polarizations, which can be converted to Stokes I, Q, U, and V parameters in post-processing.  Before each observation, both a test scan on a known pulsar (PSR~B0525+21) and a noise-diode calibration scan (for polarimetric calibration) were performed.

Dedispersed time series with DM = $461-661$\,pc\,cm$^{-3}$, in trial steps of $1$\,pc\,cm$^{-3}$, were searched using PRESTO's {\tt single\_pulse\_search.py}, which applies a matched-filter technique to look for
bursts with durations between $81.92$\,$\mu$s to $24576$\,$\mu$s (for any putative burst that only has a single peak with width $< 81.92$\,$\mu$s, the sensitivity will be degraded by a factor of a few, at most).
The resulting DM-time-S/N events were grouped into plausible astrophysical burst candidates using a custom sifting algorithm and then a dynamic spectrum of each candidate was plotted for human inspection and grading.  We found 16 bursts of astrophysical origin, and used the DSPSR package to form full-resolution, full-polarization PSRCHIVE `archive' format files for each burst.

\noindent {\it Green Bank Telescope}

On August 26, 2017, we observed \frb\ using the GBT `C-band' receiver ($4-8$\,GHz, with dual linear receptors) as part of a program of monitoring known FRB positions.  Observations were conducted with the Breakthrough Listen Digital Backend\cite{Mac17}, which allowed recording of baseband voltage data across the entire nominal 4-GHz bandwidth of the selected receiver.
Scans of a noise-diode calibration, of the flux calibrator 3C161 and of the bright pulsar PSR~B0329+54 supplemented the observations.

In post-processing, a total intensity, low-resolution filterbank data product was searched
for bursts with DM$ = 500 - 600$\,pc\,cm$^{-3}$, using trial DMs in steps of 0.1\,pc\,cm$^{-3}$ and a GPU-accelerated search package to perform the incoherent dedispersion\cite{Bar12}. We detected\cite{Gaj17} 15 bursts with S/N $> 10$.  Here we present the properties of just the two brightest GBT bursts in order to confirm the large RM observed by Arecibo and to quantify its variation in time.  A detailed analysis of all GBT detections is presented in Gajjar et al. (in prep.).
A section of raw voltage data (1.5\,s) around each detected burst was extracted and coherently
dedispersed to a nominal DM of 557.91\,pc\,cm$^{-3}$ using the DSPSR package.
Final PSRFITS format data products have time and spectral resolutions of $10.24$\,$\mu$s and $183$\,kHz, respectively.

\subsubsection*{Data analysis}

\noindent {\it Calculation of burst RMs} \\

We calibrated the burst `archives' using the PSRCHIVE utility {\tt pac} in `SingleAxis' mode.  This calibration strategy uses observations of a locally generated calibration signal (pulsed noise diode) to correct the relative gain and phase difference between the two polarization channels, under the assumption that the noise source emits equal power and has zero intrinsic phase difference in the two hands.  This calibration scheme does not correct for cross-coupling or leakage between the polarizations.  While leakage must be present at some level, the high polarization fraction, complete lack of circular polarization, and consistency of the test pulsar observations with previous work all give us confidence that calibration issues are not a significant source of error for the RM determination.  In addition, the flux density of GBT observations was calibrated using the flux calibrator.

We initially performed a brute force search for peaks in the linear polarization fraction (Extended Data Fig.~\ref{fig:LI_RM}), and discovered $\text{RM}_{\rm obs} \sim +10^5$\,rad\,m$^{-2}$ in the Arecibo data.
Each burst was corrected for Faraday rotation using the best-fit RM value for that burst.
Residual variations in the resulting PA($\lambda$) were used to refine the initial values by fitting
\be
\text{PA}(\lambda) = \text{RM}\lambda^2 + \text{PA}_{\infty}.
\ee
Subsequently, the equation
\be
\hat{L}=\exp\left\lbrace i\cdot 2(\text{RM}\lambda^2+\text{PA}_{\infty})\right\rbrace,
\ee
where $\hat{L}$ is the unit vector of the linear polarization, was used to fit the whole sample of bursts together, imposing a different RM per day and a different PA$_{\infty}$ per telescope.
The results of these fits are reported in Table~\ref{tab:bursts} and an example is shown in Fig.~\ref{fig:UQ}.

Applying the optimal RM value to each burst, we produced polarimetric profiles showing that each burst is consistent with being $\sim$100\% linearly polarized after accounting for the finite widths of the PUPPI frequency channels (Fig.~\ref{fig:bursts_zoom}; Extended Data Fig.~\ref{fig:LI_f}).
In fact, the measured Arecibo bursts are depolarized to $\sim$98\%, consistent with an uncorrected intra-channel Faraday rotation of
\be\label{eq:faraday_smearing}
\Delta\theta = \frac{{\rm RM}_{\rm obs}c^2\Delta\nu}{\nu_c^{3}},
\ee
where $c$ is the speed of light, $\Delta{\nu}$ is the channel width, and $\nu_c$ is the central channel observing frequency.  At 4.5\,GHz this corresponds to $\sim$9$^{\circ}$, and the depolarization fraction is
\be\label{eq:depol}
f_{\rm depol} = 1 - \frac{{\rm sin}(2\Delta{\theta})}{2\Delta{\theta}} = 1.6\%.
\ee

We supplemented our above analysis with a combination of RM Synthesis and {\tt RMCLEAN} (e.g. Extended Data Fig.~\ref{fig:rmclean}).
Ensuring the presence of minimal Faraday complexity is possible by integrating across the full bandwidth and taking advantage of a Fourier transform relation between the observed $\vec{L}(\lambda^2)$ values and the Faraday spectrum (the polarized brightness as a function of RM). This approach is commonly known as RM Synthesis\cite{Bre05}, and can be coupled with a deconvolution procedure ({\tt RMCLEAN}) to estimate the intrinsic Faraday spectrum\cite{Hea09}. While RM Synthesis and {\tt RMCLEAN} can have difficulty in properly reconstructing the intrinsic Faraday spectrum under certain circumstances, the spread of clean components is a reliable indicator of spectra that contain more than a single Faraday-unresolved source\cite{And15}.

At each observed frequency, we integrated Stokes Q and U values across the pulse width and normalized using Stokes I. Due to the normalization we only used frequency bins that had a Stokes I signal-to-noise ratio of at least 5. We computed a deconvolved Faraday spectrum for each burst separately, on a highly oversampled RM axis ($\delta\mathrm{RM}\approx\,10^{-4}$ of the nominal FWHM of the RM resolution element). We used a relatively small gain parameter (0.02) and terminated the deconvolution when the peak of the residual decreased to $2\sigma$ above the mean. The algorithm typically required $50-80$ iterations to converge. This combination of settings permits us to carefully consider the cumulative distribution of {\tt RMCLEAN} components along the RM axis, and thus constrain the intrinsic width of the polarized emission to $\lesssim0.1\%$ of the typical RM uncertainty. We found that this value scales with $\delta\mathrm{RM}$ because the peak of the Faraday spectrum rarely lands precisely on an individual pixel. To a high degree of confidence, there is no evidence for emission at more than one RM value, nor for a broadened (``Faraday thick'') emitting region; we therefore forego more complicated QU-fitting\cite{O'Sul12}. Results of this analysis are shown in Extended Data Table~\ref{tab:rm_synth}, and are consistent with the simplified QU-fitting results described above.

\noindent {\it Calculation of burst properties}

As in previous studies\cite{Spi16,Sch16}, a search for periodicity in the burst arrival times remains inconclusive.

Determining the exact DMs of the bursts is complicated by their changing morphology with radio frequency\cite{Spi16,Sch16}.  Measuring DM based on maximizing the peak S/N of the burst often leads to the blurring of burst structure and, in the case of \frb, an overestimation of DM.  We have thus chosen to display all bursts dedispersed to the same nominal DM from Burst \#6 (Fig.~\ref{fig:bursts_zoom} and Extended Data Fig.~\ref{fig:bursts}).
Taking advantage of the narrowness of Burst \#6, we estimated its optimal DM by minimizing its width at different DM trials.
We measured burst widths at half the maximum by fitting von Mises functions using the PSRCHIVE routine {\tt paas} (Table~\ref{tab:bursts}).
These widths correspond to the burst envelope in the case of multi-component bursts.

Flux densities of the Arecibo bursts were estimated using the radiometer equation to calculate the equivalent RMS flux density of the noise:
\be
\sigma_{\rm noise} = \frac{T_{\rm sys}}{G\sqrt{2Bt_{\rm int}}},
\ee
where $T_{\rm sys} \sim 30$\,K and $G \sim 7$\,K\,Jy$^{-1}$ are the system temperature and gain of the receiver, respectively, $B=800$\,MHz is the observing bandwidth and $t_{\rm int} = 10.24$\,$\mu$s is the sampling time.
GBT observations were instead calibrated using a flux calibrator as discussed above.
Due to the complicated spectra of the bursts, we quote average values across the frequency band (Table~\ref{tab:bursts}).

The burst dynamic spectra in Extended Data Fig.~\ref{fig:bursts} show narrow-band striations that are consistent with diffractive interstellar scintillations caused by turbulent plasma in the Milky Way.  Autocorrelation functions (ACFs) of burst spectra show three features: a very narrow feature from radiometer noise, a narrow but resolved feature corresponding to the striations, and a broad feature related to the extent of the burst across the frequency band.   The striation feature has a half width that varies from $\sim$2 to 5\,MHz from burst to burst and is comparable to the scintillation bandwidth expected from the Milky Way in the direction of \frb.   The NE2001 electron density model\cite{Cor02} provides an estimate $\tau \sim 16~\mu$s for the pulse broadening at 1 GHz. This predicts a scintillation bandwidth $\sim \nu^{4.4} / 2\pi\tau$ that ranges from 5 to 11~MHz across the 4.1 to 4.9~GHz band.
We conclude that the  measured ACFs and the NE2001 model prediction are consistent to within their uncertainties and that the narrow striations are due to Galactic scintillations.

\subsubsection*{A model for \frb's rotation measure (RM) and scattering measure (SM)}

\noindent {\it RM constraints}

The measured RM$_{\rm obs} \sim +1\times10^5\ \RMunits$ implies a source frame value
\be
{\rm RM}_{\rm src} = (1+z)^2 {\rm RM}_{\rm obs} \sim +1.4\times 10^5 \ \RMunits.
\ee

We can use the previously estimated\cite{Ten17} $\text{DM}_{\rm Host} \sim 70$--270\,pc\,cm$^{-3}$ (in the source frame) and RM$_{\rm src}$ to constrain the properties of the region in which the Faraday rotation
occurs.  In the absence of other information, we can set a constraint on the average magnetic field
along the line of sight in the Faraday region with the ratio
\be
B_{\parallel} = \frac{{\rm RM}_{\rm src}}{0.81 {\rm DM}_{\rm Host}} = [0.6\ {\rm mG}, 2.4 \ {\rm mG}].
 \ee

If only a small portion of \frb's total DM is from the highly magnetized region, the field could be much higher.

\noindent {\it SM constraints}

 The best constraint on pulse broadening comes from the measurement of the scintillation (diffraction) bandwidth
 of $\Delta\nu_{\rm d} \sim 5$~MHz at 4.5\,GHz (see above).   This implies a pulse broadening time at 1\,GHz:
  \be
 \tau(1\ {\rm GHz}) \approx (2\pi \Delta\nu_{\rm d})^{-1} \times (4.5~{\rm GHz}/1~{\rm GHz})^{4.4} = 24\ \mu s.
 \ee
 This scattering time is consistent with that expected from the Milky Way using the NE2001 model\cite{Cor02} and therefore is an upper bound on any contribution from the host galaxy.     Compared to scattering in the Milky Way,  this upper bound is below the mean trend for any of the plausible values of DM$_{\rm Host}$, especially when the correction from spherical to plane waves is taken into account\cite{Cor16b}.

 The ratio host-galaxy $\tau / \DM$ is a factor $(1+z)^2 = 1.42 $ larger in the source frame but that is still far from sufficient to account for the apparent scattering deficit compared to the Galactic $\tau$-DM relation.    Given the apparent extreme conditions of the plasma in the host galaxy, it would  not be surprising if  its turbulence properties cause a scattering deficit.  For example, scattering is reduced if the inner scale is comparable or larger than the Fresnel scale, either due to a large magnetic field or a high temperature.

\subsubsection*{Constraints on the properties of the Faraday region}

Comparison of the magnetic field and thermal energy densities enables us to constrain the density ($n_e$), electron temperature ($T_e$), and length scale ($L_{\rm RM}$) of the region responsible for the observed Faraday rotation.
We parametrize this relation with
\begin{equation}
\beta \frac{B^2}{8\pi} = 2 n_e k_B T_e,
\end{equation}
where $\beta$ is a scaling factor, $B$ is the magnetic field strength, and $k_B$ is the Boltzmann constant.  This assumes a 100\% ionized gas of pure hydrogen with temperature equilibration between protons and electrons.  Under equipartition, $\beta=1$.  In more densely magnetized regions, $\beta \ll 1$.  Field reversals will reduce the total RM, requiring a lower value of $\beta$ in order to match constraints.  The absence of free-free absorption at a frequency of $\sim$1\,GHz sets an additional constraint on the permitted parameter space.

In Extended Data Fig.~\ref{fig:dmrm}, we explore a range of physical environments.  We consider a smaller lower limit, i.e. ${\rm DM} = 1$\,pc\,cm$^{-3}$, on the dispersion measure than the previously estimated\cite{Ten17} $\text{DM}_{\rm Host} \sim 70$--270\,pc\,cm$^{-3}$, because not all of the DM may originate from the Faraday region.  Galactic HII regions typically show $| {\rm RM} | \lesssim 3 \times 10^2$\,rad\,m$^{-2}$ and weak magnetic fields\cite{Har11} with $\beta \gtrsim 1$, although calculations suggest it is possible for HII regions to achieve high RMs under some circumstances\cite{Sic17}.  Parameter space for typical HII region plasma at $T_e=10^4$\,K is almost entirely excluded, and considering a range of possible HII regions sizes and densities\cite{Hunt09} shows that these are incompatible with the $\text{DM}_{\rm Host}$ constraints.    At higher $T_e$, wide ranges of parameter space are permitted.  In the case of equipartition, we have explicit unique solutions. For $T_e=10^6$\,K, we find a density of $n_e\sim 10^2\,{\rm cm^{-3}}$ on a length scale $L_{\rm RM}\sim 1$\,pc, i.e., comparable to the upper limit on the size of the persistent source.  Higher temperature gas ($T_e=10^8$\,K) can be extended to $L_{\rm RM}\sim 100$\,pc.  For both of these solutions, the characteristic magnetic field strength is $\sim$1\,mG.

The large RM of \frb\ is similar to those seen toward massive black holes; notably, $\text{RM} \sim -5 \times 10^5$\,rad\,m$^{-2}$ is measured toward Sgr~A*, the Milky Way's central black hole,  and probes scales of $< 10^4$ Schwarzschild radii ($\sim$0.001\,pc)\cite{Bow03,Mar07}.  The constraints on $n_e$, $T_e$, and $L_{\rm RM}$ are also consistent with the environment around Sgr~A* (Extended Data Fig.~\ref{fig:dmrm}).  The high RM toward the Galactic Centre magnetar
PSR~J1745$-$2900 (Fig.~\ref{fig:DM_RM_all}), $\text{RM} = -7 \times 10^4$\,rad\,m$^{-2}$,
at a projected distance of $\sim$0.1\,pc from Sgr~A* \cite{Eat13,Sha13}, is evidence
for a dynamically organized magnetic field around Sgr~A* that extends out to the magnetar's distance\cite{Eat13}.  Notably, $\sim$4.5 years of radio monitoring of PSR~J1745$-$2900 has shown a $\sim$5\% decrease in the magnitude of the observed RM, while the DM remained constant at the $\sim$1\% level (Desvignes et al., in prep.).  This suggests large fluctuations in magnetic field strength in the Galactic Centre, on scales of roughly $10^{-5}$ parsec.

The high RM and the rich variety of other phenomena\cite{Spi14,Spi16,Sch16,Cha17,Ten17,Mar17,Sch17,Bas17} displayed by the \frb\ system suggest that the persistent radio counterpart to \frb\ could represent emission from an accreting massive black hole, with the surrounding star formation representing a circum-black-hole starburst.  Given the mass of the host galaxy, and typical scaling relationships\cite{Rei15}, the mass of the black hole would be $\sim$10$^{4-6}$\,\msun.  The observed radio brightness, compactness, and the current optical and X-ray
 non-detections\cite{Ten17,Mar17,Sch17} are compatible with such a black hole and an inefficient accretion state ($\sim$10$^{-6}$--$10^{-4}$\,$\mathrm{L_{Edd}}$, where $L_{\rm Edd}$ is the Eddington luminosity).

While models considering the presence of only a massive black hole have been proposed\cite{Vie17}, there is no observational precedent for microsecond bursts created in such environments.
Rather, the \frb\ bursts themselves could arise from a neutron star, perhaps highly magnetized and rapidly spinning, near an accreting massive black hole.  The proximity of PSR J1745$-$2900 to Sgr A* demonstrates that such a combination is
possible.  In this model, the black hole is responsible for the observed persistent source, whereas the bursts are created in the magnetosphere of the nearby neutron star\cite{Pen15}.

Alternatively, the association of \frb\ with a persistent radio source has been used to argue that the radio bursts are produced by a young magnetar powering a luminous wind nebula\cite{Met17,Kas17}.  This model is not well motivated by Galactic examples, since the most luminous (non-magnetar powered) Galactic pulsar wind nebula is only $2\times 10^{-6}$ times as luminous as the persistent source coincident with \frb, and Galactic magnetars have no detectable persistent radio wind nebulae\cite{Hes08,Rey17}.  Also, while giant flares from magnetars can produce relativistic outflows\cite{Fra99}, an upper limit on the RM from one such outburst\cite{Gae05} is 4 orders of magnitude below that observed for \frb.

Nonetheless, under the millisecond magnetar model, the properties of the persistent source constrain the putative magnetar's age to be between several years and several decades with a spin-down luminosity of $10^8$~to $10^{12}$ times higher than any local analog\cite{Met17}. Furthermore, the millisecond magnetar model predicts that the nebula magnetic field strength scales with the integrated spin-down luminosity of the magnetar\cite{Met17,Kas17}.
Extended Data Fig.~\ref{fig:dmrm} describes a range of sizes, densities, and temperatures for the Faraday-rotating medium that are consistent with Crab-like pulsar wind nebulae, known supernova remnants, and a simple model for swept-up supernova ejecta.

\subsubsection*{Data availability}

The calibrated burst data are available, upon request, from the Corresponding Author.

\subsubsection*{Code availability}

The code used to analyse the data is available at the following sites:\\
PRESTO (\url{https://github.com/scottransom/presto}),\\
PSRCHIVE (\url{http://psrchive.sourceforge.net}),\\
DSPSR (\url{http://dspsr.sourceforge.net}).

\clearpage


\clearpage


\section*{Extended Data}

\renewcommand{\baselinestretch}{1.0}
\selectfont

\noindent
Extended Data Table 1: {\bf List of $\sim$4.5\,GHz Arecibo observations used in this study.}  These are a subset of all \frb\ observations to date.

\bigskip\noindent
Extended Data Table 2: {\bf Results of RM Synthesis and {\tt RMCLEAN}.} RMs were determined by fitting a quadratic function to the peak of the deconvolved Faraday spectrum.  RM uncertainties were determined by dividing the nominal FWHM of the RM resolution element by twice the signal-to-noise ratio at the peak of the RM spectrum.  RMdisp is the second moment (dispersion) of the {\tt RMCLEAN} clean components discovered during the Faraday spectrum deconvolution. Upper limits indicate that the value scales with RM pixel size. A value of zero means that all clean components fell within the same pixel, and indicates a Faraday spectrum that is indistinguishable from being infinitely thin.

\bigskip\noindent
Extended Data Figure 1: {\bf Pulse profiles and spectra of the 16 Arecibo bursts.}
The bursts are de-dispersed to $\text{DM}=559.7$\,pc\,cm$^{-3}$ (which minimizes the width of Burst \#6) and plotted with time and frequency resolutions of $20.48$\,$\mu$s and $6.24$\,MHz, respectively.

\bigskip\noindent
Extended Data Figure 2: {\bf
Polarimetric properties of the 11 brightest bursts detected by Arecibo.}
\textbf{a}: linear polarization fraction of the bursts as a function of frequency.
A solid line shows the theoretical depolarization due to intra-channel Faraday rotation calculated using Eqs.~\ref{eq:faraday_smearing} and \ref{eq:depol}.
\textbf{b}: PA$_{\infty}$ as a function of frequency.
For both panels, values are averaged over $16$ consecutive channels.
\textbf{c}: PA$_{\infty}$ as a function of time.
A time offset is applied to each burst in order to show them consecutively.
Vertical, dashed lines divide different observing sessions.
All values in this figure have been corrected for the RM calculated with a global fit.
Grey regions in \textbf{b} and \textbf{c} indicate the 1-$\sigma$ uncertainty around the PA value from the global fit.

\bigskip\noindent
Extended Data Figure 3: {\bf
Linear polarization fraction of the bursts as a function of RM.}
Different colours represent different observing sessions (see legend).
A grey line indicates the average RM yielding the largest polarization fraction in the first observing session.

\bigskip\noindent
Extended Data Figure 4: {\bf Example RM Synthesis and {\tt RMCLEAN} for Burst~\#8.}  The relevant RM range is shown for Burst~\#8, after RM Synthesis (dashed line) and {\tt RMCLEAN} (solid line), as described in the text. Only two clean components (red circles) were required to reach convergence in the deconvolution algorithm (at $102,679.5$ and $102,679.75\,\mathrm{rad\,m}^{-2}$; cf. the peak of the final deconvolved Faraday spectrum at $102,679.65\,\mathrm{rad\,m}^{-2}$). For all bursts, the RM Synthesis and {\tt RMCLEAN} steps demonstrate an extremely thin and single-peaked Faraday spectrum.

\bigskip\noindent
Extended Data Figure 5: {\bf
RM and PA$_{\infty}$ values of the different bursts.}
Coloured, 1-$\sigma$ error bars represent individual bursts, with central values highlighted by black dots.
Horizontal grey regions are values obtained from a global fit.
Values used in the figure are reported in Table~\ref{tab:bursts}.

\bigskip\noindent
Extended Data Figure 6: {\bf Physical constraints from source parameters.}  Parameter space for electron density ($n_e$) and length scale ($L_{\rm RM}$) of the Faraday region
for three different temperature regimes, $T_e=[10^4, 10^6, 10^8]$ K.
The shaded red region indicates parameter space excluded by optical depth considerations ($\tau_{ff} > 5$).
The solid black line gives the maximum DM$_{\rm Host}$ permitted, while the shaded grey region shows the DM
down to 1 ${\rm pc\, cm^{-3}}$.
The solid blue line gives RM$_{\rm src}$.  The shaded blue region gives the range $10^{-4} <= \beta <= 1$.
The intersection of grey and blue regions outside of the red region are physically permitted.
The arrows indicates the upper limits on the sizes of the persistent source (left) and the star-forming region (right), respectively\cite{Mar17,Ten17}.
The parallel dashed lines represent fits to a range of galactic and extragalactic HII regions\cite{Hunt09}.  The parallel dotted lines represent the evolution of 1 and 10 \msun\ of ejecta evolving up to 1000 years at a velocity of $10^4 {\rm\, km\,s^{-1}}$ in the blast-wave phase following a supernova\cite{McKee95}. The filled downwards triangle and diamond are for the supernova remnants Cas~A\cite{Orl16} and SN~1987A\cite{McC16}, respectively.  The filled circle represents the mean density and diameter of the Crab Nebula, whereas the filled square represents the characteristic density and length scale of a dense filament in the Crab Nebula\cite{Dav85}.  The star indicates the density at the Bondi radius of Sgr~A*\cite{Qua99}.
\clearpage

\captionsetup[table]{name=Extended Data Table}
\setcounter{table}{0}
\begin{table}
\centering
\caption{{\bf List of $\sim$4.5\,GHz Arecibo observations used in this study.}  These are a subset of all \frb\ observations to date.} \label{tab:observations}
\centerline{\includegraphics{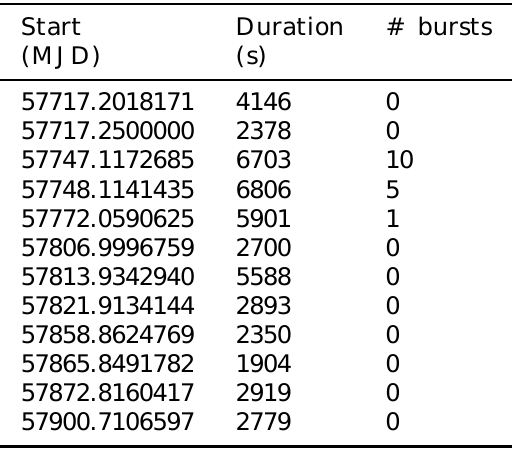}}
\end{table}

\clearpage

\begin{table}
\centering
\caption{{\bf Results of RM Synthesis and {\tt RMCLEAN}.} RMs were determined by fitting a quadratic function to the peak of the deconvolved Faraday spectrum.  RM uncertainties were determined by dividing the nominal FWHM of the RM resolution element by twice the signal-to-noise ratio at the peak of the RM spectrum.  RMdisp is the second moment (dispersion) of the {\tt RMCLEAN} clean components discovered during the Faraday spectrum deconvolution. Upper limits indicate that the value scales with RM pixel size. A value of zero means that all clean components fell within the same pixel, and indicates a Faraday spectrum that is indistinguishable from being infinitely thin.} \label{tab:rm_synth}
\centerline{\includegraphics{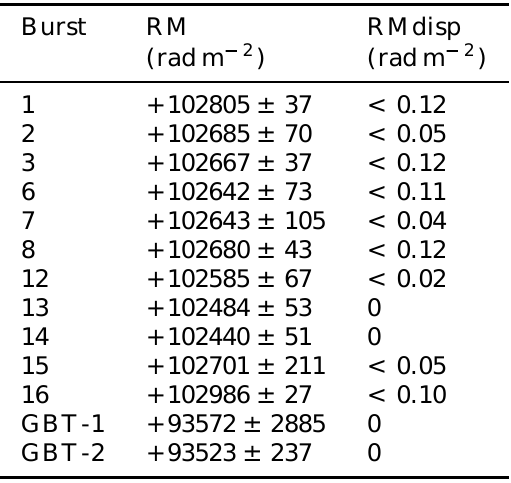}}
\end{table}

\clearpage

\setcounter{figure}{0}

\begin{figure}
\centerline{\includegraphics{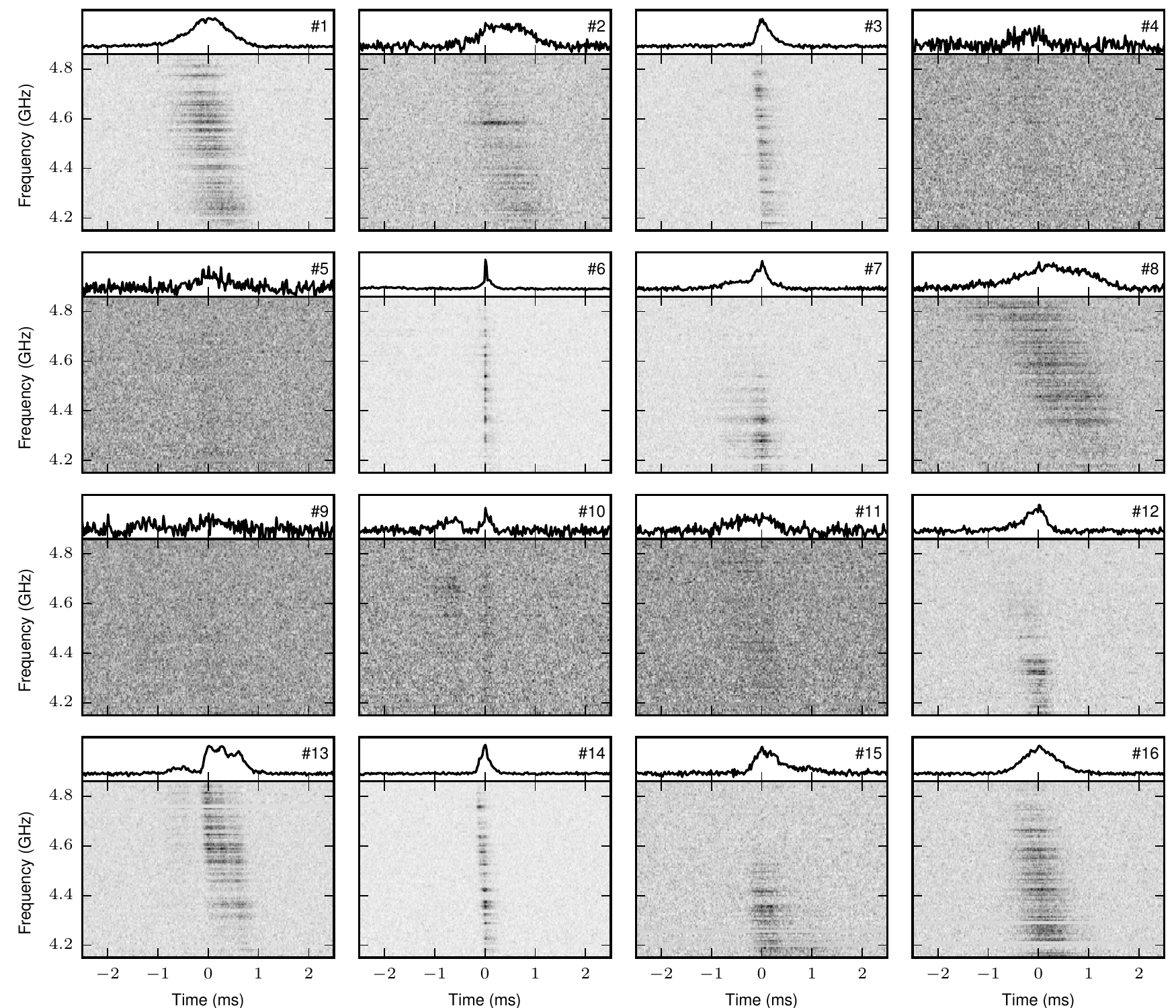}}
\caption{{\bf Pulse profiles and spectra of the 16 Arecibo bursts.}
The bursts are de-dispersed to $\text{DM}=559.7$\,pc\,cm$^{-3}$ (which minimizes the width of Burst \#6) and plotted with time and frequency resolutions of $20.48$\,$\mu$s and $6.24$\,MHz, respectively.
\label{fig:bursts}}
\end{figure}

\clearpage

\begin{figure}
\centerline{\includegraphics{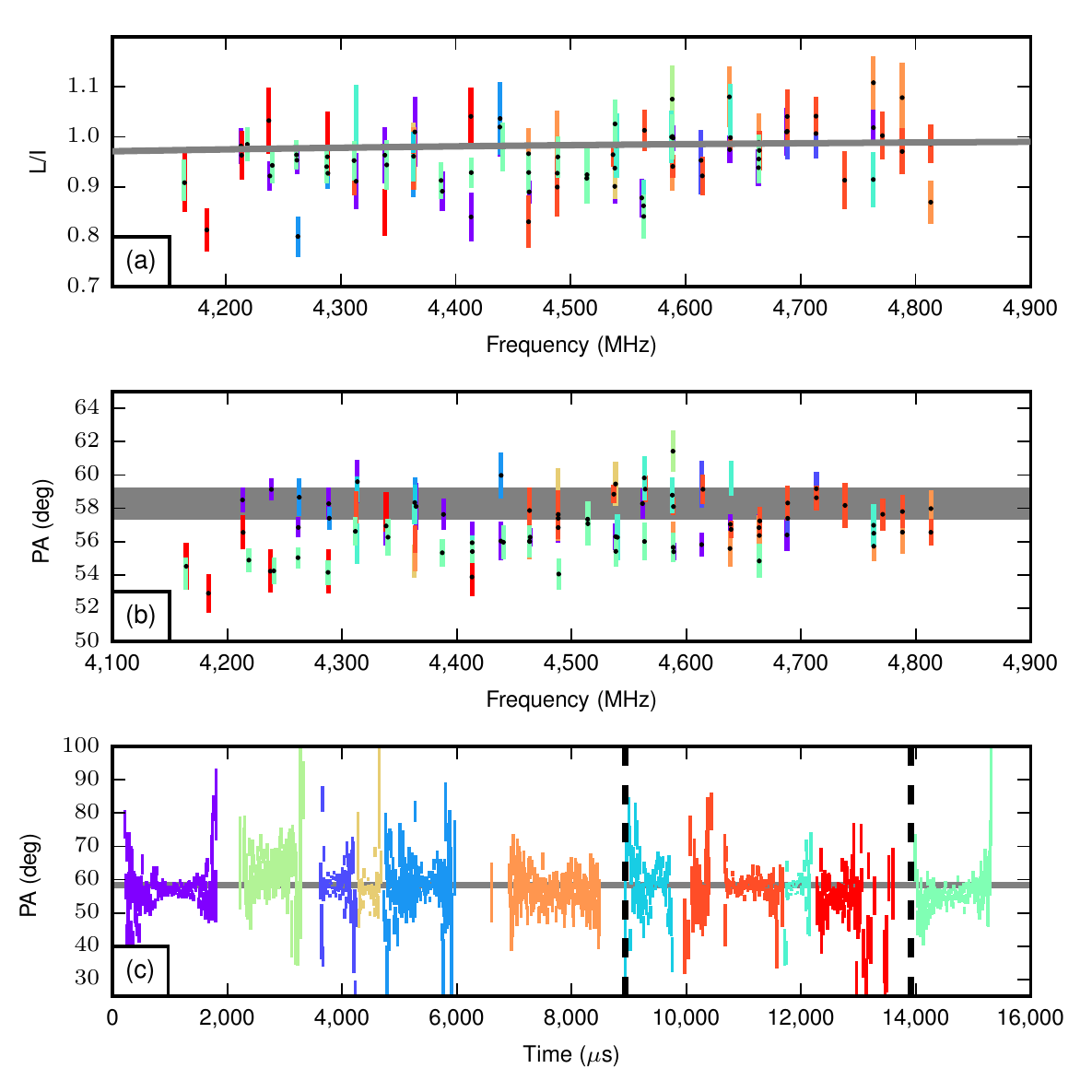}}
\caption{{\bf
Polarimetric properties of the 11 brightest bursts detected by Arecibo.}
\textbf{a}: linear polarization fraction of the bursts as a function of frequency.
A solid line shows the theoretical depolarization due to intra-channel Faraday rotation calculated using Eqs.~\ref{eq:faraday_smearing} and \ref{eq:depol}.
\textbf{b}: PA$_{\infty}$ as a function of frequency.
For both panels, values are averaged over $16$ consecutive channels.
\textbf{c}: PA$_{\infty}$ as a function of time.
A time offset is applied to each burst in order to show them consecutively.
Vertical, dashed lines divide different observing sessions.
All values in this figure have been corrected for the RM calculated with a global fit.
Grey regions in \textbf{b} and \textbf{c} indicate the 1-$\sigma$ uncertainty around the PA value from the global fit.
\label{fig:LI_f}}
\end{figure}

\clearpage

\begin{figure}
\centerline{\includegraphics{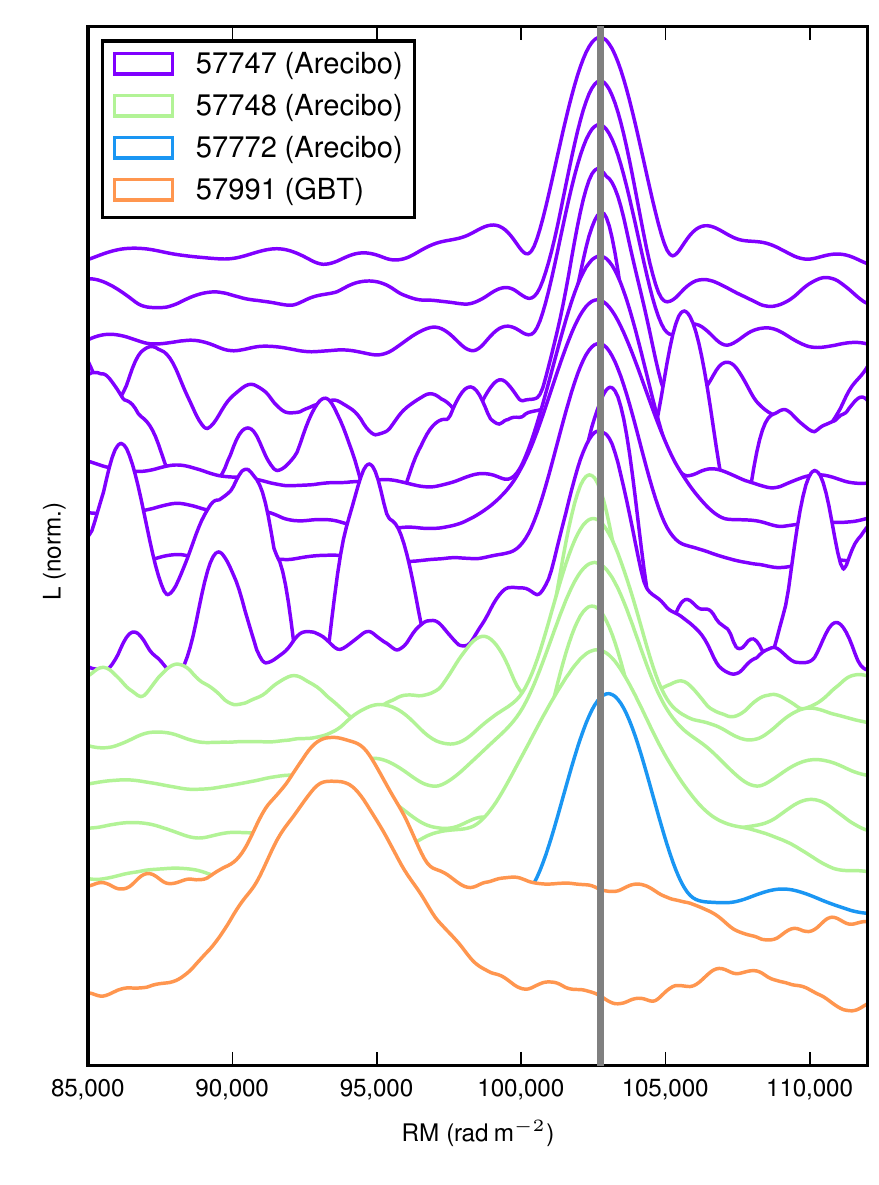}}
\caption{{\bf
Linear polarization fraction of the bursts as a function of RM.}
Different colours represent different observing sessions (see legend).
A grey line indicates the average RM yielding the largest polarization fraction in the first observing session.
\label{fig:LI_RM}}
\end{figure}

\clearpage

\begin{figure}
\centerline{\includegraphics{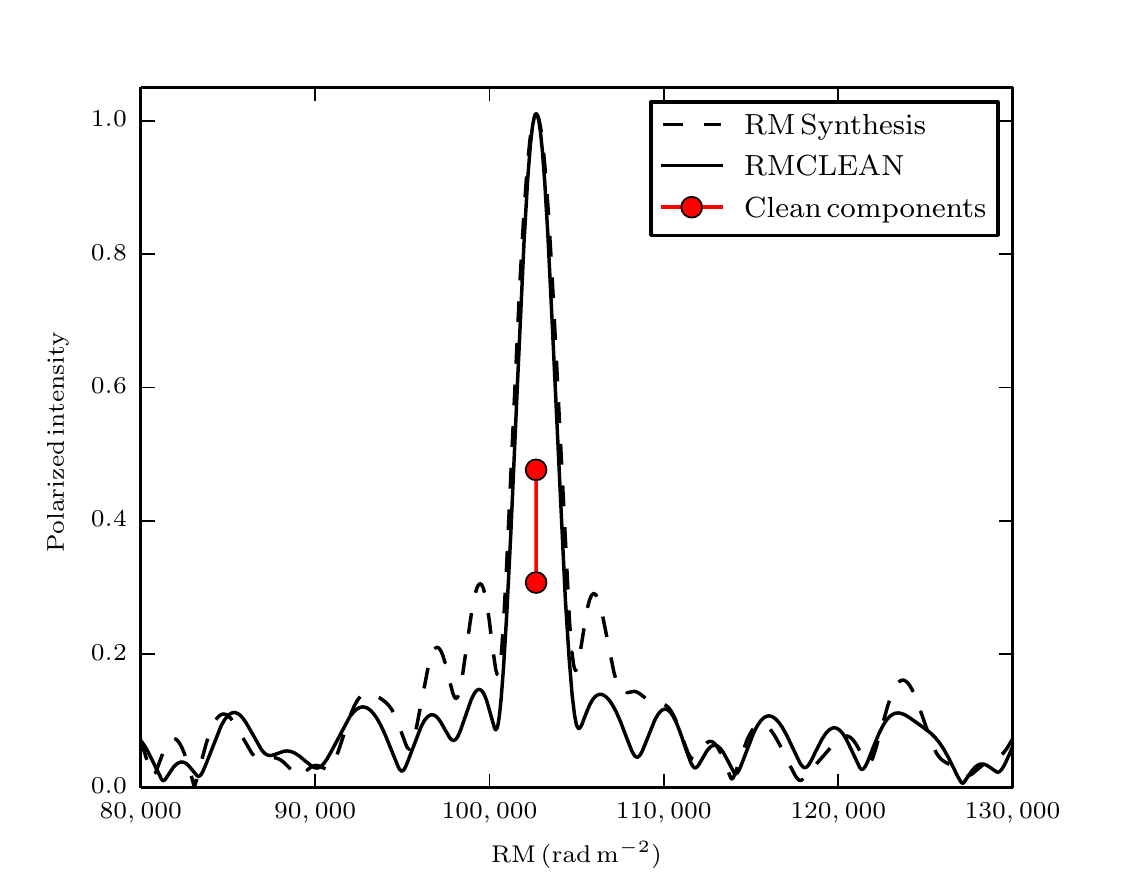}}
\caption{{\bf Example RM Synthesis and {\tt RMCLEAN} for Burst~\#8.}  The relevant RM range is shown for Burst~\#8, after RM Synthesis (dashed line) and {\tt RMCLEAN} (solid line), as described in the text. Only two clean components (red circles) were required to reach convergence in the deconvolution algorithm (at $102,679.5$ and $102,679.75\,\mathrm{rad\,m}^{-2}$; cf. the peak of the final deconvolved Faraday spectrum at $102,679.65\,\mathrm{rad\,m}^{-2}$). For all bursts, the RM Synthesis and {\tt RMCLEAN} steps demonstrate an extremely thin and single-peaked Faraday spectrum.
\label{fig:rmclean}}
\end{figure}

\clearpage

\begin{figure}
\centerline{\includegraphics{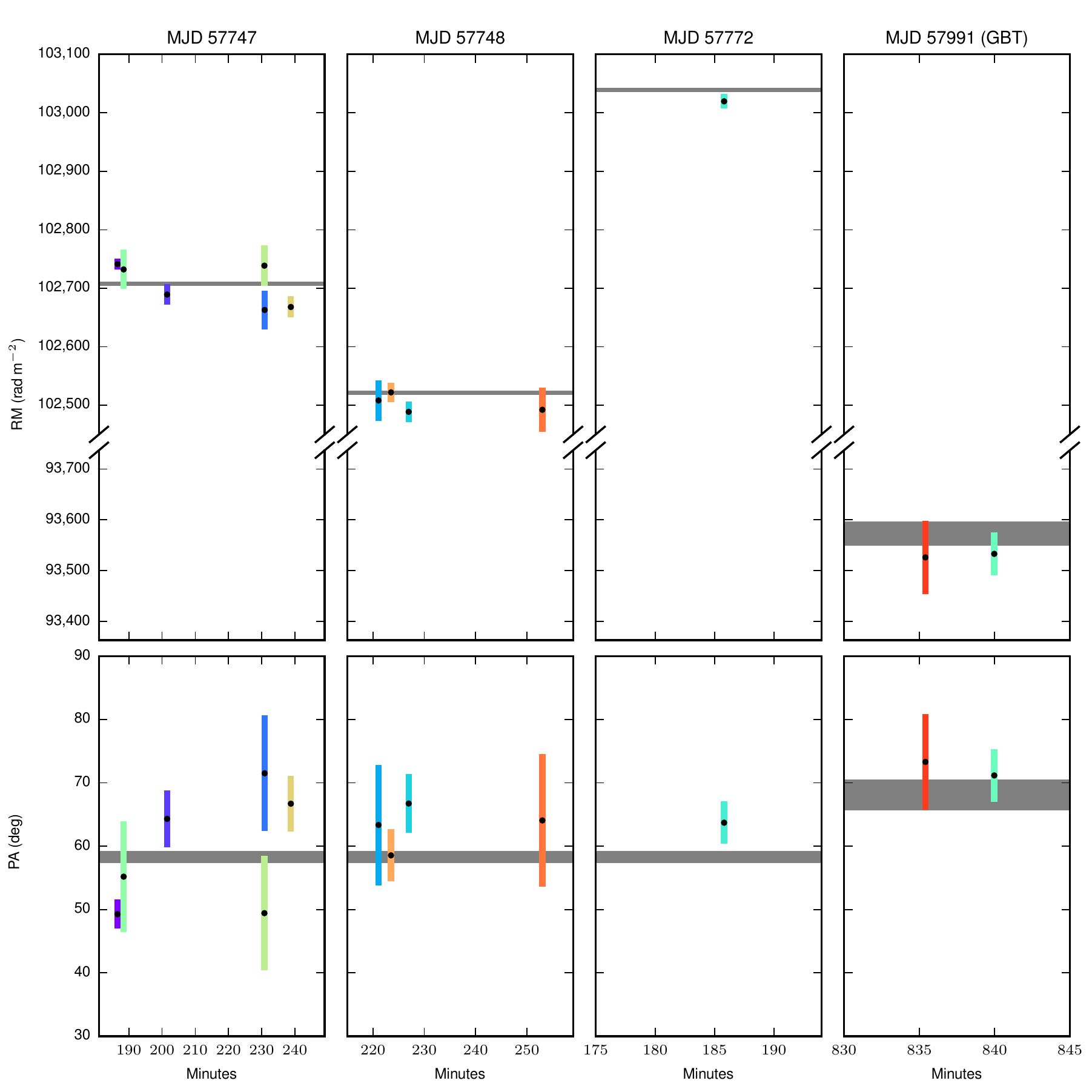}}
\caption{{\bf
RM and PA$_{\infty}$ values of the different bursts.}
Coloured, 1-$\sigma$ error bars represent individual bursts, with central values highlighted by black dots.
Horizontal grey regions are values obtained from a global fit.
Values used in the figure are reported in Table~\ref{tab:bursts}.
\label{fig:RMevolution}}
\end{figure}

\clearpage

\begin{figure}
\centerline{\includegraphics{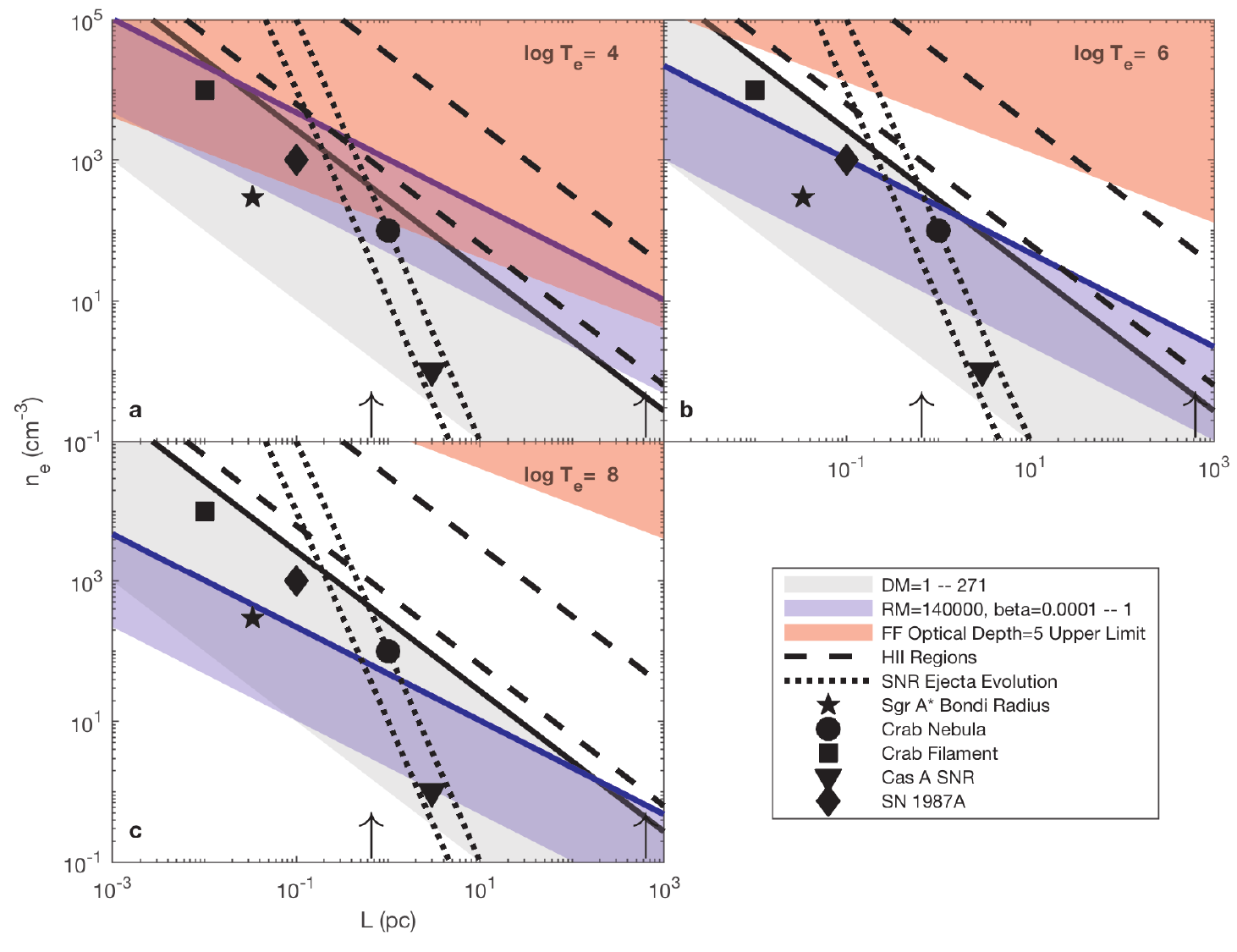}}
\caption{{\bf Physical constraints from source parameters.}  Parameter space for electron density ($n_e$) and length scale ($L_{\rm RM}$) of the Faraday region
for three different temperature regimes, $T_e=[10^4, 10^6, 10^8]$ K.
The shaded red region indicates parameter space excluded by optical depth considerations ($\tau_{ff} > 5$).
The solid black line gives the maximum DM$_{\rm Host}$ permitted, while the shaded grey region shows the DM
down to 1 ${\rm pc\, cm^{-3}}$.
The solid blue line gives RM$_{\rm src}$.  The shaded blue region gives the range $10^{-4} <= \beta <= 1$.
The intersection of grey and blue regions outside of the red region are physically permitted.
The arrows indicates the upper limits on the sizes of the persistent source (left) and the star-forming region (right), respectively\cite{Mar17,Ten17}.
The parallel dashed lines represent fits to a range of galactic and extragalactic HII regions\cite{Hunt09}.  The parallel dotted lines represent the evolution of 1 and 10 \msun\ of ejecta evolving up to 1000 years at a velocity of $10^4 {\rm\, km\,s^{-1}}$ in the blast-wave phase following a supernova\cite{McKee95}. The filled downwards triangle and diamond are for the supernova remnants Cas~A\cite{Orl16} and SN~1987A\cite{McC16}, respectively.  The filled circle represents the mean density and diameter of the Crab Nebula, whereas the filled square represents the characteristic density and length scale of a dense filament in the Crab Nebula\cite{Dav85}.  The star indicates the density at the Bondi radius of Sgr~A*\cite{Qua99}.  \label{fig:dmrm}}
\end{figure}

\end{document}